\documentclass[12pt]{iopart}
\usepackage[colorlinks=true,linkcolor=blue,urlcolor=blue,citecolor=blue,linktocpage=true]{hyperref}
\usepackage{amssymb, braket,graphicx}
\expandafter\let\csname equation*\endcsname\relax
\expandafter\let\csname endequation*\endcsname\relax
\usepackage{amsmath}

\def\be{\begin{equation}}
\def\ee{\end{equation}}
\newcommand{\m}{{\rm m}}
\newcommand{\rss}{{\rm ss}}

\newcommand{\Hext}{H_{\rm BWD}^{\rm ext}}
\newcommand{\Yext}{Y_{\rm BWD}^{\rm ext}}
\newcommand{\Hextre}{H_{\rm BWD,recon.}^{\rm ext}}

\begin{document}
\title[Probing Primordial SGWB with Multi-band Astrophysical Foreground Cleaning]{Probing Primordial Stochastic Gravitational Wave Background with Multi-band Astrophysical Foreground Cleaning }
\author{Zhen Pan}
\ead{zpan@perimeterinstitute.ca}
\address{Perimeter Institute for Theoretical Physics, Ontario, N2L 2Y5, Canada}
\author{Huan Yang}
\ead{hyang@perimeterinstitute.ca}
\address{Perimeter Institute for Theoretical Physics, Ontario, N2L 2Y5, Canada}
\address{University of Guelph, Guelph, Ontario N2L 3G1, Canada}

\date{\today}
\begin{abstract}
The primordial stochastic gravitational wave background (SGWB) carries first-hand messages of early-universe physics,
possibly including effects from inflation, preheating, cosmic strings, electroweak symmetry breaking, and etc. However, the astrophysical foreground  from compact binaries may mask the SGWB, introducing difficulties in detecting the signal and measuring it accurately.
In this paper, we propose a foreground cleaning method taking advantage of gravitational wave observations in other frequency bands. We apply this method to probing the SGWB with space-borne gravitational wave detectors, such as
the  Laser Interferometer Space Antenna (LISA). We find that the spectral density of the LISA-band astrophysical foreground from compact binaries (black holes and neutron stars)
can be predicted with percent-level accuracy assuming $10$-years' observations of third-generation GW detectors,
e.g., Cosmic Explorer. While this multi-band method does not apply to binary white dwarfs (BWDs) which usually merger
before entering the frequency band of ground-based detectors, we limit our foreground
cleaning to frequency higher than $\sim 5$ mHz, where all galactic BWDs can be individually resolved by LISA and
the shape of the spectral density of the foreground from extragalactic BWDs can be reconstructed and/or modeled with certain uncertainties. After the foreground cleaning, LISA's sensitivity to the primordial SGWB will be substantially improved
for either two LISA constellations where SGWB can be measured by cross correlating their outputs  or
only one constellation with three spacecrafts where SGWB can be measured by constrasting the responses of a signal channel and
a null channel.
\end{abstract}
\maketitle

\section{Introduction}
Primordial stochastic gravitational wave background (SGWB) has been conjectured to arise from
various fundamental physical processes from the early universe \cite{Caprini2018,Christensen2019},
including the inflationary origin
\cite{Guth1981,Linde1982,Starobinskii1979,Turner1997,Easther2006,Easther2007,Barnaby2012,Cook2012},
cosmic strings
\cite{Kibble1976,Vilenkin1981,Hogan1984,Caldwell1992,Vilenkin1994,Damour2000,Damour2001,Jeannerot2003,Damour2005,Sakellariadou2006,Sakellariadou2007,Siemens2007} and
first-order phase transition due to electroweak symmetry breaking \cite{Turner1990,Turner1992,Kosowsky1992,Kosowsky1992b,Kamionkowski1994,Caprini2008,Iso2017,Cutting2018}.
Therefore measuring primordial SGWB at different frequencies will provide  important information
 to understand our universe before recombination \cite{Kamionkowski1997,Seljak1997,Hobbs2010,Bender1998,DECIGO2017,LVC2017,PhysRevX.6.011035}.
However, the total SGWB also contains contribution from astrophysical foreground of gravitational waves (GWs)  from unresolved compact binaries \cite{Farmer2003,Marassi2011,Rosado2011,Zhu2011,Wu2012,Zhu2013,LVC2017,LVC2018b,Mauro2020}, including  binary white dwarfs (BWDs), binary black holes (BBHs), binary neutron stars (BNSs) and possibly black hole-neutron star binaries (BHNSs). Removing the influence  by the astrophysical foreground would be an essential step towards the measurements of primordial SGWB.

Inspired by recent discussions about the benefits of multi-band GW observations
\cite{Sesana2016,Barausse2016,Vitale2016,Tinto2016,Sesana2017,Wong2018,Isoyama2018,Carson2019},
we propose a multi-band foreground cleaning method and apply it to measuring
the primordial SGWB in the LISA band \cite{Bartolo2016}. The third-generation GW detectors, e.g.
Cosmic Explorer (CE) \cite{Abbott2017CE} and Einstein Telescope (ET) \cite{Punturo:2010zz},
are expected to detect almost all BBH and BNS mergers in our universe \cite{Regimbau2017,LVC2019}.
With data from these ground-based detectors, we can reconstruct the underlying distribution
of the BBH/BNS population and derive their contribution to the astrophysical foreground in the LISA band. In particular,
we find that the astrophysical foreground from compact binaries  can be predicted with percent-level accuracy with the CE running for $10$ years.
After removing this predicted astrophysical foreground from the LISA data, it can be shown that LISA's sensitivity to the primordial
SGWB will be substantially enhanced.

In this work, we use BBHs as the proxy of compact binaries, while the astrophysical foreground
sourced by BNSs and BHNSs can be cleaned in the same way. This multi-band foreground cleaning method does not apply
for the galactic BWDs, because BWDs merge at a much lower frequency and never enter the  band of ground-based detectors.
Therefore we conservatively confine our analysis to a higher frequency band ($f\gtrsim 5$ mHz)
where the galactic BWDs can be completely resolved by LISA \cite{Lamberts2019} (see \cite{Cutler2006, Adams2014}
for details of galactic BWDs subtraction).
In this paper we do not consider another frequent astrophysical GW source in the universe, supernovae.
As shown in Ref.~\cite{Buonanno2005}, most GW emission from type II supernovae  is at frequencies
higher than 1 Hz and  the contribution to astrophysical foreground at LISA band is very small
(e.g., comparing to the BBHs contribution), even considering rather optimistic model with very anisotropic Emission.
The GW emission from type Ia supernovae of BWD mergers also turns out to be much weaker
than in the chirp phase \cite{Loren2009,Dan2011}.

This paper is organized as follows. In Section~\ref{sec:BBH}, we outline the basic formulas for calculating the spectral
density of stochastic GWs from BBHs. In Section~\ref{sec:reconstruction}, we explain how to reconstruct the distribution
of BBHs from merger events detected by ground based detectors and quantify the uncertainty in estimating the BBH foreground.
In Section~\ref{sec:BWD} and \ref{sec:primordial}, we show the LISA sensitivity to the  extraglatic BWD foreground
and to the primordial SGWB will be substantially improved with
the multi-band cleaning of BBH foreground  assuming two constellations in orbit. The influence of possible eccentric BBHs is briefly discussed in Section~\ref{sec:discussion}. In \ref{sec:onebyone}, we discuss  an alternative approach of BBH foreground cleaning, the reconstruction
of extragalactic BWD foreground and the foreground cleaning with  a single space-borne constellation.
We use the geometrical units $G=c=1$ and
assume a flat $\Lambda$CDM cosmology with $H_0= 70$ km/s/Mpc, $\Omega_{\Lambda} = 0.7$ and $\Omega_{\rm m} = 0.3$.

\section{Stochastic GWs from BBHs}
\label{sec:BBH}
Assuming the SGWB is isotropic, unpolarized and stationary, we can define its spectral density $H(f)$ as (our definition is
different from that of Ref.~\cite{Allen1999} by a factor $8\pi$),
\be\label{eq:sd}
\braket{h^*_A(f,\hat\Omega) h_B(f',\hat\Omega')}
= \frac{1}{2}\frac{\delta_{\hat\Omega,\hat\Omega'}}{4\pi}\delta_{AB} \delta(f-f') H(f)\ ,
\ee
with $h_{A,B}(f,\hat\Omega)$ being the waveform of gravitational waves coming from direction $\hat\Omega$ with polarization state
$A,B\in\{+,\times\}$ written in the Fourier domain. The spectral density $H(f)$ is related to the energy density of the SGWB by
\be\label{eq:rhogw}
\rho_{\rm GW} = \frac{\pi}{2}\int_0^\infty f^2 H(f) df\ ,
\ee
which in turn relates to the energy fraction of GWs in a logarithmic frequency bin $\Omega_{\rm GW}(f)$ by
\be\label{eq:omega}
\Omega_{\rm GW}(f) := \frac{1}{\rho_{\rm crit}} \frac{d\rho_{\rm GW}}{d \ln f} = \frac{4\pi^2}{3H_0^2}|f|^3 H(|f|)\ ,
\ee
where $\rho_{\rm crit}:=3 H_0^2/8\pi$ is critical energy density to close the universe.

The energy density of GWs averaged
over all inspiral binaries in different directions $\hat \Omega$ and a period of time $[-T/2, T/2]$ is  \cite{Phinney2001}
\be\label{eq:rho_a}
\begin{aligned}
  \hat \rho_{\rm GW}
  &= \frac{1}{T} \sum_i \int_{-T/2}^{T/2} S_i(t) dt \\
  &= \frac{1}{T} \sum_i \int_{-T/2}^{T/2}  \frac{1}{16\pi}(\dot h_+^2 + \dot h_\times^2)_i dt \\
  &= \frac{\pi}{2}\frac{1}{T} \sum_i \int_{f_-^i}^{f_+^i}  f^2\left(|h_+(f)|^2+|h_\times(f)|^2 \right)_i df,
\end{aligned}
\ee
with index $i$ running over all BBHs in the universe,
$S_i(t)$ being the energy flux of GWs emitted by the $i$-th BBH,
dots denoting time derivative
and $f^i_\pm = f^i|_{t=\pm\frac{T}{2}}$ being the GW frequency of $i$-th BBH at $t=\pm T/2$.
According to Eqs.(\ref{eq:rhogw}) and (\ref{eq:rho_a}), we find the spectral density of the BBH
foreground averaged over time period $[-T/2, T/2]$ is
\be\label{eq:HA_oneone}
\hat H_{\rm A}(f) = \frac{1}{T}\sum_i  \left(|h_+(f)|^2+|h_\times(f)|^2 \right)_i \Theta_T^i(f) \ ,
\ee
where $\Theta_T^i(f) = 1$ if $f\in [f^i_-, f^i_+]$ and $\Theta_T^i(f) = 0$ otherwise.

To obtain the mean value of $\hat H_{\rm A}(f)$, we first consider a sample of BBHs with same redshift $z_s$,
same chirp mass $M_s$ and merger rate $\dot N_s$. For this sample, all BBHs evolve along the same frequency-time curve $f(t-t^i_{\rm merger})$, i.e.,
$\frac{1}{T}\braket{\sum_s  \Theta_T^s(f)}$ is independent of frequency $f$ (as long as $f$ is lower than the merger frequency)
and is equal to $\dot N_s$.
Therefore, we have
\be
  \braket{\hat H_{\rm A}(f)}_s
  = \braket{\left(|h_+(f)|^2+|h_\times(f)|^2 \right)_s }\times  \dot N_s\ .
\ee
Now consider BBHs in the real universe, with a merger rate density $R(z)$
(number of mergers per comoving volume per unit of cosmic time local to the event)
and the chirp mass distribution $p(M_c)$, we have the merger rate
$\dot N_s(z,M_c) dz dM_c =  \frac{R(z)}{1+z}dV_c(z) p(M_c) d M_c$ and the mean value of $\hat H_{\rm A}(f)$ \cite{Phinney2001}
\be\label{eq:HA_mean}
\braket{\hat H_{\rm A}(f)} =  \int_0^\infty\int_{M_{\rm c,min}}^{M_{\rm c,max}} \braket{\left(|h_+(f)|^2+|h_\times(f)|^2 \right)_s}\dot N_s dz dM_c\ ,
\ee
where $ dV_{\rm c}(z) = 4\pi r^2(z)/H(z) dz$ is the comoving volume element,
with $r(z) = \int_0^z dz/H(z)$ being the comoving radial distance and $H(z)$ being the Hubble expansion rate at redshift $z$.
In the quasi-circular approximation, the waveform in the LISA band is \cite{Cutler1994}
\be\label{eq:waveform}
\begin{aligned}
  h_+(f) &= \frac{1+\cos^2\imath}{2}\sqrt{\frac{5}{24}} \frac{(G\mathcal M_z)^{5/6}f^{-7/6}}{\pi^{2/3}D_{\rm L}} e^{-i\Psi(f)}\ , \\
  h_\times(f) &= i\cos\imath \sqrt{\frac{5}{24}} \frac{(G\mathcal M_z)^{5/6}f^{-7/6}}{\pi^{2/3}D_{\rm L}} e^{-i\Psi(f)}\ ,
\end{aligned}
\ee
where $\imath$ is the inclination angle of the binary orbital direction with respect to the line of sight to observers on the earth,
$\mathcal M_z=(1+z)M_c$ is the redshifted chirp mass,
$D_{\rm L}$ is the luminosity distance and $\Psi(f)$ is the wave phase.
Plugging Eq.~(\ref{eq:waveform}) into Eq.~(\ref{eq:HA_mean}), we find
\be\label{eq:HA_avg}
\braket{\hat H_{\rm A}(f)} = \dot N_{\rm O} \frac{f^{-7/3}}{6\pi^{4/3}} \int_0^\infty P(\zeta)\zeta^2 d\zeta\ ,
\ee
where
\[ \dot N_{\rm O} = \int_0^\infty\int_{M_{\rm c,min}}^{M_{\rm c,max}}  \dot N_s(z,M_c) dzdM_c \ ,\]
is the merger rate seen in the observer's frame,
$\zeta = (G\mathcal M_z)^{5/6}/D_{\rm L}$ is the amplitude and $P(\zeta)$ is the corresponding probability distribution.

Physically  Eqs.~(\ref{eq:HA_oneone}) and (\ref{eq:HA_avg}) display two different perspectives in understanding the astrophysical foreground. The former describes an event-based approach: the foreground consists of GWs from all unresolved (by the LISA) inspiral binaries, each of which will enter the ground detector band at a later time; therefore the foreground may be
estimated by summing up contribution from events later identified by CE or ET.  Eq. (\ref{eq:HA_avg}) states that the (ensemble average of) foreground can be obtained from the statistical distribution of the BBHs, which may also be measured precisely
by ground-based detectors. While both approaches are equivalent given infinite detector running time and accuracy,  we will show later that the distribution-based approach is much more efficient than the event-to-event subtraction given a finite detector running time, say $10$ years.
In the following sections, we will adopt the distribution-based approach
and briefly discuss the application of Eq.~(\ref{eq:HA_oneone}) in an  event-to-event subtraction in the \ref{sec:onebyone}.

To describe the underlying distribution of the BBH mergers, we need to specify the local merger rate density $R(z)$ (number of mergers per comoving volume per unit of cosmic time) and the mass distribution $p(m_1, m_2)$.
As a fiducial model, we assume $R(z) = R_0 e^{-(z/10)^2}$ with $R_0 = 65\ {\rm Gpc}^{-3} {\rm yr}^{-1}$  \cite{LVC2018},
and
\be
p(m_1, m_2) \propto \frac{1}{m_1(m_1-m_{\rm min})} \ ,
\ee
for $m_{\rm min} \leq m_2 \leq m_1 \leq m_{\rm max}$, with $m_{\rm min} = 5 M_\odot$ and $m_{\rm max} = 42 M_\odot$.
Combining the mass distribution $p(m_1, m_2)$ with the merge rate density $R(z)$, it is straightforward to infer
the merger rate $\dot N_{\rm O}$ in the observer's frame and  the underlying distribution function
$P(\zeta)$ (see Fig.~\ref{fig:kde}). Given $R(z)$ and $p(m_1,m_2)$, we  generate many samples of  BBH mergers,
then reconstruct the distribution function $P(\zeta)$  assuming a number of  merger events are recorded by ground-based detectors.

\section{Distribution Reconstruction}
\label{sec:reconstruction}
We model the Fourier-domain waveform $h_A(f)$ of BBH mergers with the PhenomB waveform \cite{Ajith2011} which depends on $7$ parameters:
redshifted chirp mass $\mathcal M_z$, redshifted total mass $M_z$, luminosity distance $D_{\rm L}$, effective spin
$\chi$, merger time $t_0$, merger phase $\varphi_0$ and inclination angle $\imath$.
The measured strain $h(f)$ is related to $h_A(f)$ by
\be
h(f) = h_+(f) F^+(f;\theta, \phi, \psi) + h_\times(f) F^\times(f;\theta, \phi, \psi)\ ,
\ee
where $F^{+,\times}(f;\theta, \phi, \psi)$ are the detector response functions which depend on
the  sky location $\theta, \phi$ and the polarization angle $\psi$.
As an example, we consider a network with three detectors (assuming CE sensitivity \cite{Abbott2017CE}) located in Australia, China and US respectively \cite{Zhao2018}, with their
locations and arm orientations specified in Table~\ref{tab}, where the orientation is the angle between the bisector of two detector
arms and the local west-to-east direction.
\begin{table}
\begin{center}
  \caption{\label{tab} Location and orientation of the three ground-based detectors considered in this work.}
  \begin{tabular}{ l | c  c  c }
     &  latitude & longitude &  orientation \\
    \hline
    Detector 1 & $32^\circ$ S & $115^\circ$ E & $135^\circ$ \\
    Detector 2 & $38^\circ$ N & $104^\circ$ E & $90^\circ$ \\
    Detector 3 & $31^\circ$ N & $90^\circ$ W  & $27^\circ$\\
    \hline
    \hline
  \end{tabular}
\end{center}
\end{table}

For each event, we estimate the parameter uncertainties using the Fisher matrix \cite{Cutler1994},
\be\label{eq:fish}
F_{\alpha\beta} = \sum_{d=1}^{3} F^d_{\alpha\beta}
= \sum_{d=1}^{3} 4\int_0^\infty \frac{\Re[h_{d,\alpha}(f)h_{d,\beta}^*(f)]}{P_{\rm n, CE}(f)} df\ ,
\ee
where $\Re$ denotes the real part, $h_d(f)$ is the strain in detector $d$, $h_{d,\alpha}$ is the derivative with respect to
parameter $\alpha$, and $P_{\rm n,CE}(f)$ is the noise spectral density of detectors (Fig.~1 in Ref.~\cite{Abbott2017CE}).
The 1-$\sigma$ uncertainty of parameter $\alpha$ is given by $\sigma_\alpha = \sqrt{(F^{-1})_{\alpha\alpha}}$.
During an observational period $T_{\rm D}$, we observe $N_{\rm O}$ mergers together with their best-estimated parameters
$\{\zeta_i\}$ ($i = 1,...,N_{\rm O}$), where $\zeta_{i}$ is sampled from a Gaussian distribution with mean value
$\zeta_i^{(\rm true)}$ and standard deviation $\sqrt{\zeta_{,\alpha}(F^{-1})_{\alpha\beta} \zeta_{,\beta}}$ with
$\zeta_{,\alpha}$ being the derivative of $\zeta$ over model parameter $\alpha$.

\begin{figure*}
\includegraphics[scale=0.55]{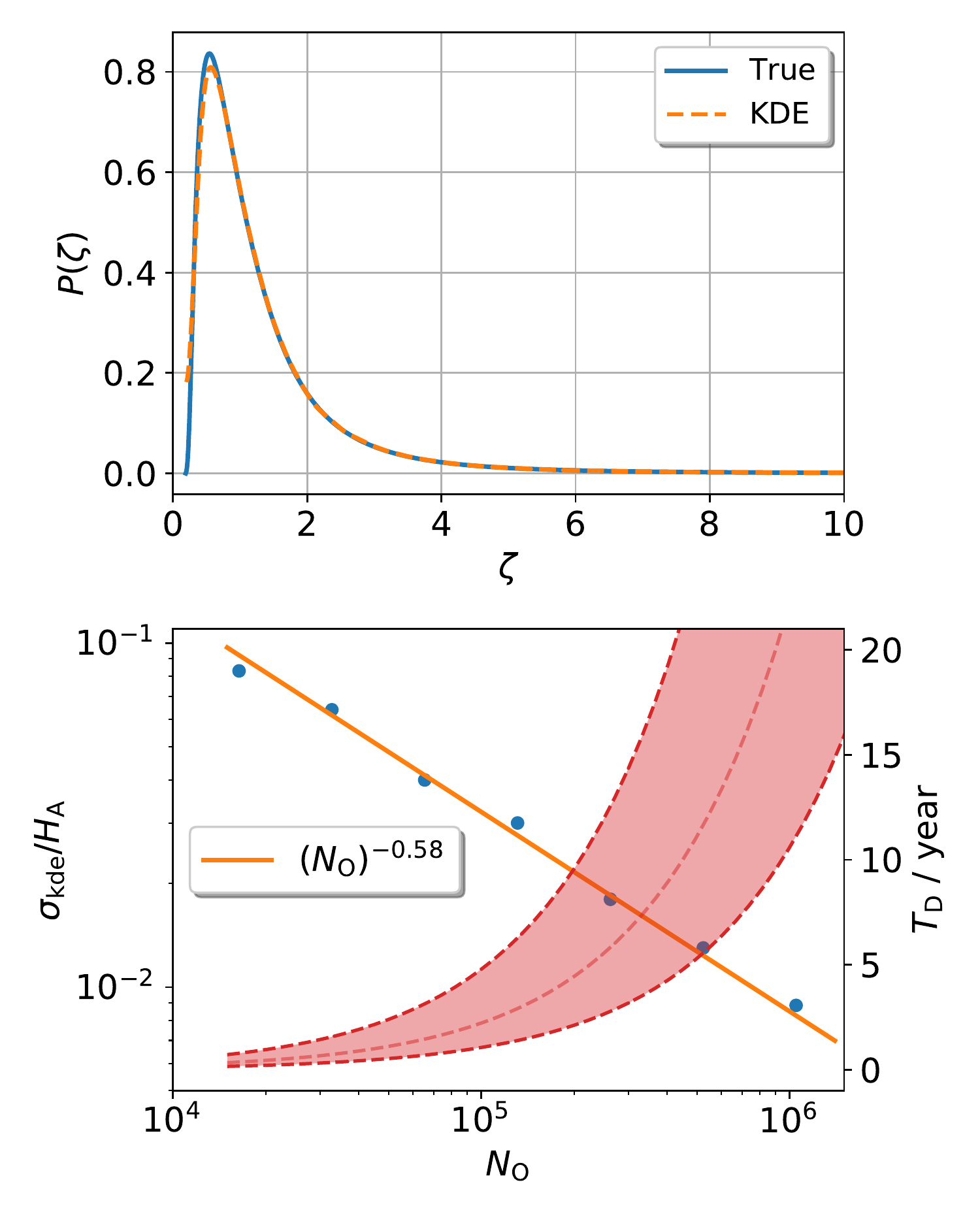}
\caption{\label{fig:kde} Upper panel:the solid line is the underlying distribution of amplitude $\zeta$ defined following Eq.~(\ref{eq:HA_avg}),
and the dashed line is the KDE reconstruction from $4.5\times 10^5$ merger events detected by the detector network, where $\zeta$ is
shown in units of $M_\odot^{5/6}  {\rm Gpc}^{-1}$.
Lower panel:the dots denote the fractional deviation
of KDE reconstructed SGWB spectral density and the solid line is
a power-law fit $\sigma_{\rm kde}/H_{\rm A}\propto N_{\rm O}^{-0.58}$.
The shadowed area
denotes the uncertainty in the amount of mergers we expect the ground-based detectors to record in a given running time $T_{\rm D}$, where the thin/dashed line in the center is the fiducial model we use in this work and the two thick/dashed lines correspond to
merger rate density $R_0 = 65^{+75}_{-34}\ {\rm Gpc}^{-3} {\rm yr}^{-1}$  \cite{LVC2018}.}
\end{figure*}

With a sample of $\{\zeta_{i}\}$, we can estimate the underlying distribution $P(\zeta)$ using the kernel density estimator (KDE).
We make use of the FFTKDE module from Python package KDEpy and determine the estimator bandwidth using
 Silverman's rule of thumb \footnote{\url{https://kdepy.readthedocs.io/en/latest/introduction.html}}.
 In Figure~\ref{fig:kde}, we show the KDE reconstructed distribution function $P(\zeta)$
 from a sample of $4.5\times 10^5$ data points. The underlying distribution is reconstructed to a good precision except
 in the range of small $\zeta$ \footnote{The KDE estimator is smoothed over its bandwidth, therefore it does not capture
 variations with length scale much shorter than the bandwidth.}.

To quantify the performance of the KDE reconstruction, we generate $100$ realizations of $N_{\rm O}$ BBH mergers,
``observe" each merger with the detector network and reconstruct $P(\zeta)$ in each realization.
With the reconstructed $P(\zeta)$,  we then calculate the spectral density
$H_{\rm kde}(f)$ using Eq.~(\ref{eq:HA_avg}). In Fig.~\ref{fig:kde}, we show the fractional deviation
$\sigma_{\rm kde}/H_{\rm A} := \sqrt{\braket{(H_{\rm kde}-H_{\rm A})^2}}/H_{\rm A}$
as a function of the total number of mergers $N_{\rm O}$. We find that the fractional bias scales as $\propto N_{\rm O}^{-0.58}$,
being $\sim 1.3\%$ for $N_{\rm O}= 4.5\times 10^5$ which is roughly the number of BBH merger events in
$10$ years.

Of course the KDE reconstruction uncertainty depends on not only the number of mergers $N_{\rm O}$
but also the detector sensitivity. The KDE is essentially an inverse-variance weighting, a method of aggregating many random variables to minimize the variance of the weighted average. Given a sample of observations $\{x_i\pm \sigma_i\} (i=1,...,N_{\rm O})$, their inverse-variance weighted average and the corresponding standard deviation are
$ \hat x = \frac{\sum_i x_i/\sigma_i^2}{\sum_i1/\sigma_i^2}$ and $\quad \sqrt{\frac{1}{\sum_i 1/\sigma_i^2}}:=\frac{\bar\sigma}{\sqrt{N_{\rm O}}}$, respectively.
In the context of detecting GW events, $\bar\sigma\propto\sigma_i\propto (\rm SNR_i)^{-1}$ is  proportional to
the detector strain sensitivity $P_{\rm n, CE}^{1/2}$ and the square root of number of detectors $N_{\rm d}^{1/2}$. Therefore the KDE reconstruction uncertainty
can be estimated via the scaling $\sigma_{\rm kde}/H_{\rm A}\propto N_{\rm d}^{1/2} P_{\rm n, CE}^{1/2}$ for different detector configurations and different detector numbers assumed.

\section{Estimating the Extragalactic BWD Foreground}
\label{sec:BWD}
Starting from the mHz range the galactic BWDs will be resolved by LISA, so that the main sources of SGWs are extragalactic BWDs \cite{Farmer2003} and compact binaries (BBHs, BNSs, and BHNSs) \footnote{The gravitational memory produced in compact binary mergers \cite{yang2018testing} is much weaker than these sources in the inspiraling stage.}.
As the compact binary foreground can be estimated with ground-based GW detections and cleaned accordingly, let us examine the detection of SGWs from extragalactic BWDs in the absence of primordial waves.

For simplicity, let us consider two concentric LISA detectors with output $s_d(t) = h_d(t) + n_d(t)$ ($d= 1,2$), with $h_d$ and $n_d$ denoting the gravitational strain and the intrinsic noise in detector $d$, respectively.
The case with a single detector is discussed in the \ref{sec:onedetector}.
The SGWs signal can be measured by
cross-correlating outputs  from two detector because the detector noises $n_1$ and $n_2$ are not correlated, while the GW signals $h_1$ and $h_2$ are correlated. In the Fourier domain,
the cross-correlation estimator is written as \cite{Allen1999}
\be\label{eq:estimator}
\hat X = \int_{-\infty}^{\infty} df \int_{-\infty}^{\infty} df' \delta_T(f-f') s^*_1(f) s_2(f') Q(f) \ ,
\ee
where $\delta_T(f):=\sin(\pi f T)/(\pi f)$ is a finite-time approximation to the $\delta$-function, $T$ is the running time of the two detectors and $Q(f)$ is a filter function. For later convenience, we define
\be
\begin{aligned}
  \braket{ H_x(f)}_a^b&:=\int_a^b H_x(f)\mathcal R_{12}(f) Q(f) df\ ,\\
  \braket{ H_x(f)}_{[a,b]}^{[b,c]}&:= \braket{ H_x(f)}_a^b- \braket{ H_x(f)}_b^c\ ,
\end{aligned}
\ee
where
\[
\mathcal R_{12}(f) = \int \frac{d\hat\Omega}{4\pi} e^{i2\pi f(\vec x_1-\vec x_2)\cdot\hat\Omega} \sum_A F_1^{A*}(\hat\Omega,f)F_2^{A}(\hat\Omega,f)\ , \]
is the overlap reduction function of the  detectors, with $\vec x_{1,2}$ being the detector locations.
Here we assume two concentric LISA detectors which form a hexagonal pattern \cite{Cornish2001}.
The mean value and variance of estimator $\hat X$ turn out to be \cite{Allen1999}
\be\label{eq:X_stat}
\begin{aligned}
  \braket{\hat X} &= T\braket{H(f)}_0^{\infty} \ ,\\
  \sigma_X^2 &\approx \frac{T}{2}\int_0^{\infty} P_{\rm n}^2(f) |Q(f)|^2 df\ ,
\end{aligned}
\ee
where
$P_{\rm n}(f)$
is the detector noise spectral density \cite{Robson2019}.
In the ideal case of zero foreground, we can  extract the primordial signal directly using the estimator (\ref{eq:X_stat}).
With the optimal filter
$ Q(f) = H(f)\mathcal R_{12}^*(f)/P_{\rm n}^2(f)$, we obtain the maximized signal-to-noise ratio (SNR)
\be
{\rm SNR}_{\rm ideal} = \sqrt{2T \int_0^{\infty}  \frac{H^2(f) }{P_{\rm n}^2(f)}|\mathcal R_{12}(f)|^2 df\ . }
\ee

The presence of astrophysical foreground of BBHs and galactic BWDs makes the problem more complicated.
In the LISA band, the foreground is dominated by the GW emission from galactic BWDs,
of which high-frequency binaries can be completely resolved and subtracted in the LISA mission time \cite{Lamberts2019}.
And we need to design an estimator with the BBHs foreground subtracted using the KDE reconstruction.
The spectral density $H_{\rm A}(f)$ of the BBH foreground is  known as a power law, while the spectral density of galactic BWD  foreground depends on their orbital distributions. In the following discussion, we
will confine our analysis to the frequency range $f\geq f_{\min} = 5$  mHz, where galactic BWD foreground
can be cleaned up and $H(f)\simeq H_{\rm BWD}^{\rm ext}(f)+H_{\rm A}(f)$ to a good approximation.

The spectral density of the astrophysical foreground from compact binaries $H_{\rm A}(f)$ can be estimated with $H_{\rm kde}(f)$
elaborated in the previous section. From Eq.~(\ref{eq:X_stat}), we define an
estimator of  the extragalactic BWD foreground as
\be\label{eq:est_Y}
  \hat Y= \hat X -T\braket{H_{\rm kde}(f)}_{f_{\rm min}}^{f_{\rm max}} \ ,
\ee
with expectation value and variance
\be
\begin{aligned}
  \braket{\hat Y}
  &= \Yext + T\braket{H_{\rm A}(f)-H_{\rm kde}(f)}_{f_{\rm min}}^{f_{\rm max}} \ ,\\
  \sigma_Y^2
  &=  \frac{T}{2}\int_{f_{\rm min}}^{f_{\rm max}} P_{\rm n}^2(f) |Q(f)|^2 df\ ,
\end{aligned}
\ee
where $\Yext = T\braket{\Hext(f)}_{f_{\rm min}}^{f_{\rm max}}$ and $f_{\rm max} = 1$ Hz.
Therefore we have $Y_{\rm BWD}^{\rm ext} = \hat Y \pm \sigma_Y(\rm stat.)\pm  \sigma_{\rm A} (\rm syst.)$,
where $\sigma_Y\propto \sqrt{T}$ is the statistical uncertainty due to detector noise and
$\sigma_{\rm A} =  T\braket{\sigma_{\rm kde} H_{\rm A}(f)}_{f_{\rm min}}^{f_{\rm max}}$
is the systematic bias due to the limited accuracy of the foreground measurement. Consequently,
the SNR of estimator $\hat Y$, ${\rm SNR}_{\hat Y}=\Yext/\sqrt{\sigma_Y^2+\sigma_{\rm A}^2}$, scales as $\sqrt{T}$ for small $T$, and saturates at $\Yext/\sigma_{\rm A}$ in the large $T$ limit.

Without the multi-band cleaning, the influence of BBH foreground may be removed by using its frequency dependence.
For later convenience, we first define a binned estimator
\be
  \hat X_{\mathcal C} := \int_{f\in \mathcal C} df \int_{f'\in\mathcal C} df' \delta_T(f-f') s^*_1(f) s_2(f') Q(f)\ ,
\ee
and also define  an estimator $\hat Z$:
\be\label{eq:est_Z}
\hat Z = \hat X_{\mathcal C_1}-\hat X_{\mathcal C_2}\ ,
\ee
where $\mathcal C_1=[f_{\rm min}, f_*], \mathcal C_2=[f_*,f_{\rm max}]$, and $f_*$ is determined by the constraint that  the influence of astrophysical foreground is removed
$\braket{H_{\rm A}(f)}_{\mathcal C_1}^{\mathcal C_2} = 0$.
The mean value and the variance of  $\hat Z$ are
\begin{equation}
  \begin{aligned}
    \braket{\hat Z} &=T\braket{\Hext(f)}_{\mathcal C_1}^{\mathcal C_2}   \ , \\
    \sigma_Z^2 &= \frac{T}{2}\int_{f_{\rm min}}^{f_{\rm max}} P_{\rm n}^2(f) |Q(f)|^2 df \ .
  \end{aligned}
\end{equation}

\begin{figure}
\includegraphics[scale=0.65]{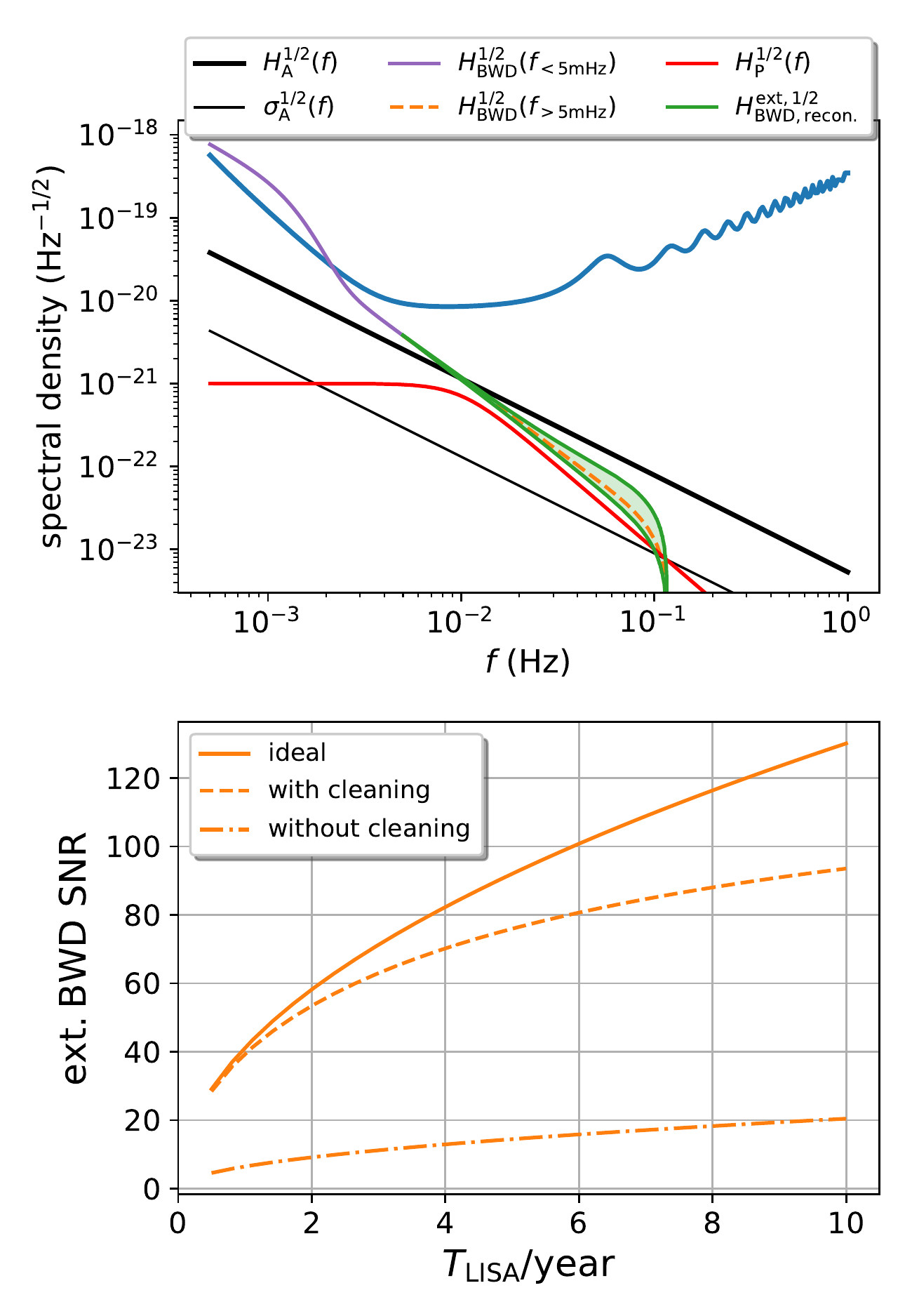}
\caption{\label{fig:bwd} Measuring the SGWs from extragalactic BWDs $\Hext(f)$.
Upper panel: the spectral densities of effective LISA detector noise $S_{\rm n}(f) $ (blue),
BBH foreground $H_{\rm A}(f)$ (thick black) and residual foreground $\sigma_{\rm A}(f)$ after cleaning (thin black),
unresolved low-frequency/high-frequency BWD foreground $H_{\rm BWD}(f)$ (purple/orange) \cite{Farmer2003,Cornish2017,Lamberts2019},
and reconstructed high-frequency BWD foreground $\Hext(f)$ (green shadow), and a primordial SGWB $H_{\rm P}(f)$ from \cite{Caprini2009} (red).
Lower panel: SNRs of different estimators for the high-frequency extragalactic BWD foreground, where the BBH foreground cleaning
increases the detection sensitivity by a factor $4\sim 7$. We also examined less optimistic scenarios of foreground cleaning with $2$ CE-like detectors, and we find the SNR (dashed line) decreases by maximally $\sim 15\%$. }
\end{figure}

In the upper panel Fig.~\ref{fig:bwd}, we present the spectral densities of LISA detector noise, stochastic GW foreground
from various sources,  and the residual foreground ($\sigma_{\rm kde}/H_{\rm A} = 1.3\%$) of compact binaries.
In the lower panel, we show the SNRs of
extragalactic BWD SGWs with  spectral density $\Hext(f)$  \cite{Farmer2003, Lamberts2019}.
Notice that without the compact binary foreground cleaning, the BWD background can still be computed by utilizing
the spectral shape of the foreground [Eq.~(\ref{eq:est_Z})], albeit with much worse sensitivity.
Such comparison is shown by the two dashed lines in the figure,
where the foreground cleaning [Eq.~(\ref{eq:est_Y})] increases the detection sensitivity by a factor of $4\sim 7$. In the same figure, the solid line depicts  the SNR of an ideal case assuming there was no foreground contamination from compact binaries.

With the presence of primordial SGWB, which is well motivated from various early universe processes,
the above analysis can be interpreted as a measurement of a combined signal of the extragalactic BWD foreground
and the primordial SGWB.

\section{Estimating the Primordial SGWB}
\label{sec:primordial}

\begin{figure}
  \includegraphics[scale=0.65]{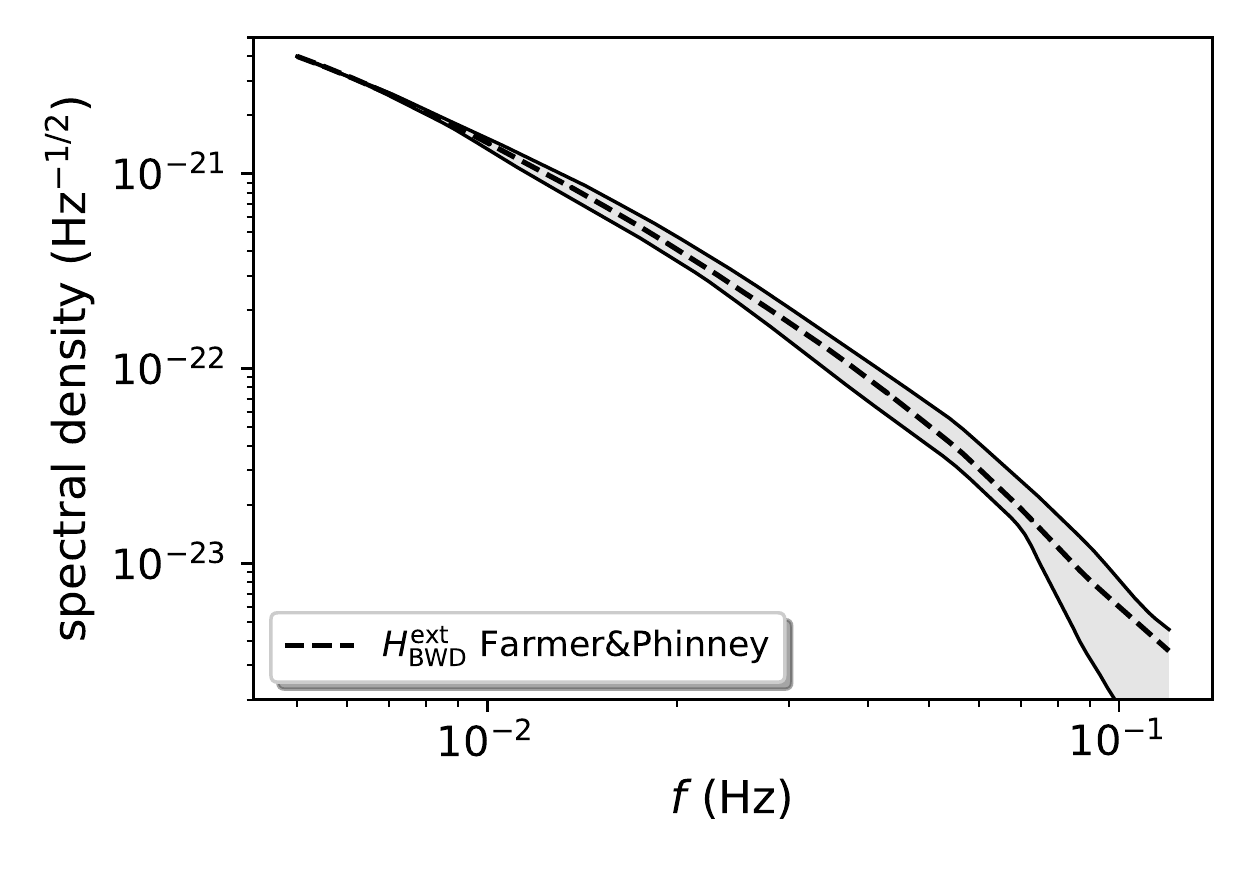}
\includegraphics[scale=0.65]{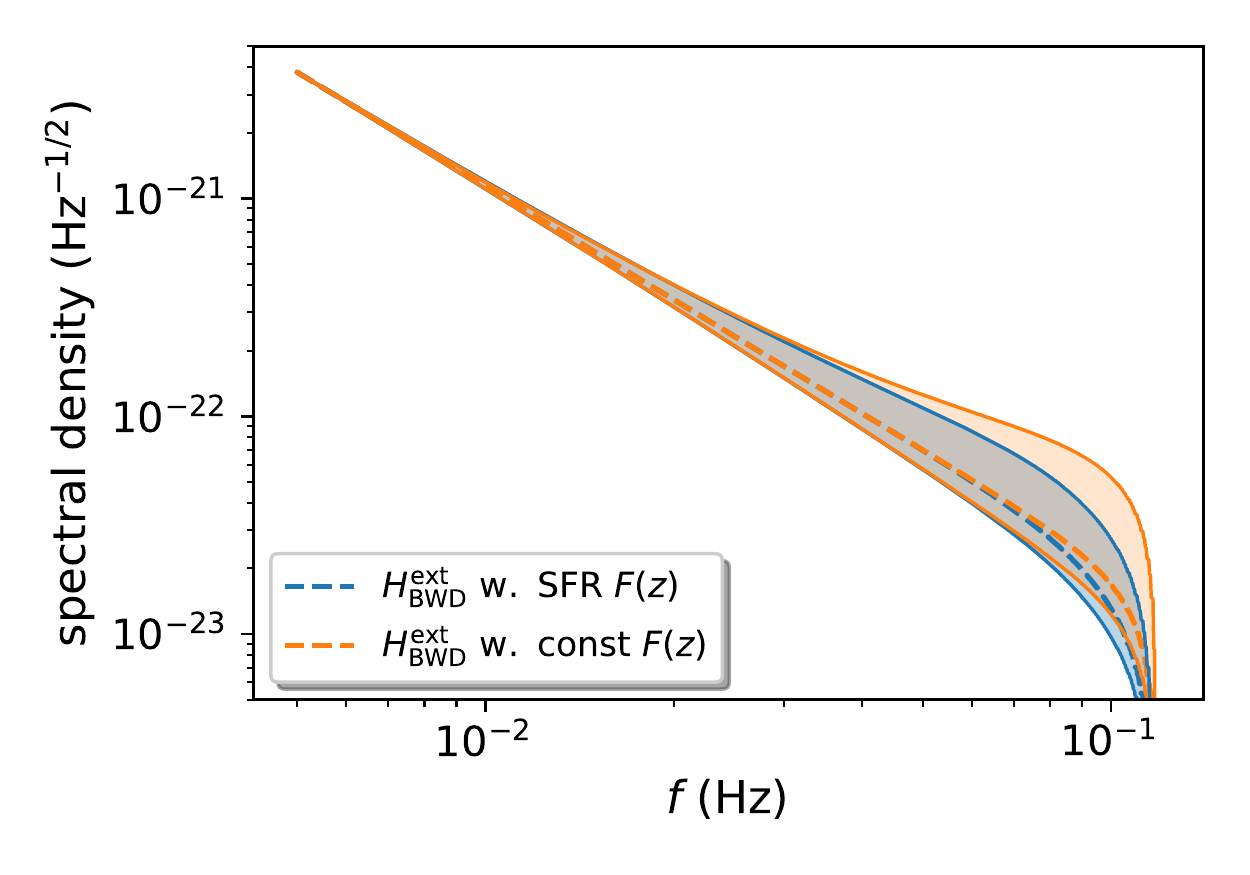}
\caption{\label{fig:Hext} The shape of spectral  density $\Hext(f)$ can be
modeled (left panel) or
reconstructed from resolvable galactic BWDs (right panel), where we have normalized
the amplitude of $\Hext(f)$ at $f=5$ mHz.
In the left panel, we have used
the variation from the optimistic foreground to the pessimistic foreground calculated in Ref.~\cite{Farmer2003}
to quantify the shape uncertainty. As shown in the right panel,
the reconstructed $\Hext(f)$ is
insensitive to the formation rate $\mathcal F(z)$ of high-frequency BWDs.
The blue line is the result of setting $\mathcal F(z)$ equal to the observed star formation rate \cite{Cole2001},
and the orange line is the result of constant $\mathcal F(z)$ (see \ref{sec:Hext} for calculation details).
The blue and orange shadows are the corresponding statistic uncertainties of reconstruction
due to limited number of resolvable galactic BWDs.
}
\end{figure}

The discussion in the previous section provides a way to  constrain $H_{\rm P}(f) + \Hext(f)$.
In order to measure the primordial component $H_{\rm P}(f)$ to probe the early universe, the shape of $\Hext(f)$ must be known with certain precision. Unlike BBHs,
BWDs of various masses and types may merge in the LISA band so that the mass distribution may have explicit frequency dependence.
The resulting stochastic background deviates from simple power laws at high frequencies (see Fig.~\ref{fig:bwd}).
A detailed theoretical study can be found in Ref.~\cite{Farmer2003}, which shows that there is a large uncertainty in the
amplitude of the extragalactic BWD foreground, while the shape is insensitive to the astrophysical uncertainties.
In this work, we propose to remove the extragalactic BWD foreground via its  frequency dependence, i.e., spectral shape, \
which can be either modeled as in Ref.~\cite{Farmer2003} or reconstructed from the individually resolved galactic BWDs
shown in Fig.~\ref{fig:Hext} and in  \ref{sec:Hext}. We find the shape uncertainties calculated from the above two methods
are comparable and display similar frequency dependence. It is promising to test the
(in)consistency of the two in the LISA era.

With the shape information of $\Hext(f)$, we can remove the extragalactic BWD foreground using its frequency dependence.
Similar to estimator $\hat Y$, we can define estimator
\be
\begin{aligned}
  \hat U& =  \hat X_{\mathcal C_1}-\hat X_{\mathcal C_2} -
  T  \braket{H_{\rm kde}}_{\mathcal C_1}^{\mathcal C_2}\ ,
\end{aligned}
\ee
where $\mathcal C_1 = [f_{\rm min}, f_*], \mathcal C_2= [f_*,f_{\rm max}]$,  $f_*$ is determined by the constraint that  the influence of extragalactic BWD foreground is removed
$\braket{\Hextre}_{\mathcal C_1}^{\mathcal C_2} = 0$.
The mean value and the variance of $\hat U$ are
\be
\begin{aligned}
\braket{\hat U}
&= T\braket{H_{\rm P} + \Hext + H_{\rm A}-H_{\rm kde}}_{\mathcal C_1}^{\mathcal C_2} \ , \\
\sigma_U^2  &= \frac{T}{2}\int_{f_{\rm min}}^{f_{\rm max}} P_{\rm n}^2(f) |Q(f)|^2 df \ ,
\end{aligned}
\ee
Similar to estimator $\hat Y$, we have $U_{\rm P}:=T\braket{H_{\rm P}}_{\mathcal C_1}^{\mathcal C_2}
= \hat U \pm \sigma_U(\rm stat.)\pm\sigma_{\rm BWD}(\rm syst.) \pm \sigma_{\rm A}(\rm syst.)$,
where $\sigma_U\propto \sqrt{T}$ is the statistical uncertainty due to detector noise,
and two systematic terms $\sigma_{\rm BWD}$ and $\sigma_{\rm A}$ are
the uncertainty of $T\braket{\Hextre}_{\mathcal C_1}^{\mathcal C_2}$ (Fig.~\ref{fig:Hext})
and $T\braket{\sigma_{\rm kde} H_{\rm A}}_{\mathcal C_1}^{\mathcal C_2}$ (Fig.~\ref{fig:kde}), respectively.

Similar to estimator $\hat Z$, both the influence of the extragalactic BWD foreground and the BBH foreground can
be removed via their frequency dependence without multi-band foreground cleaning. We can define estimator
\be
\hat V = \left(\hat X_{\mathcal C_1}-\hat X_{\mathcal C_2}\right)-\left(\hat X_{\mathcal C_3}-\hat X_{\mathcal C_4}\right)\ ,
\ee
where $\mathcal C_{1,..,4}$ are 4 non-overlapped frequency bins and are determined by the constraint that
both the extragalactic BWD foreground and the BBH foreground are removed. The mean value and the variance of $\hat V$
are
\be
\begin{aligned}
  \braket{\hat V}
  &= T\braket{H_{\rm P} + \Hext}_{\mathcal C_1}^{\mathcal C_2} - T\braket{H_{\rm P} + \Hext}_{\mathcal C_3}^{\mathcal C_4}\ ,\\
  \sigma_V^2
  &= \frac{T}{2}\int_{f_{\rm min}}^{f_{\rm max}} P_{\rm n}^2(f) |Q(f)|^2 df \ ,
\end{aligned}
\ee
We have $V_{\rm P} := T\braket{H_{\rm P}}_{\mathcal C_1}^{\mathcal C_2} - T\braket{H_{\rm P}}_{\mathcal C_3}^{\mathcal C_4}
=\hat V \pm \sigma_V(\rm stat.)\pm \sigma_{\rm BWD}(syst.) $, where $\sigma_V$ is the uncertainty due to
detector noise and $\sigma_{\rm BWD}$ is the uncertainty of $T\braket{\Hext}_{\mathcal C_1}^{\mathcal C_2} - T\braket{ \Hext}_{\mathcal C_3}^{\mathcal C_4}$.

\begin{figure}
\includegraphics[scale=0.65]{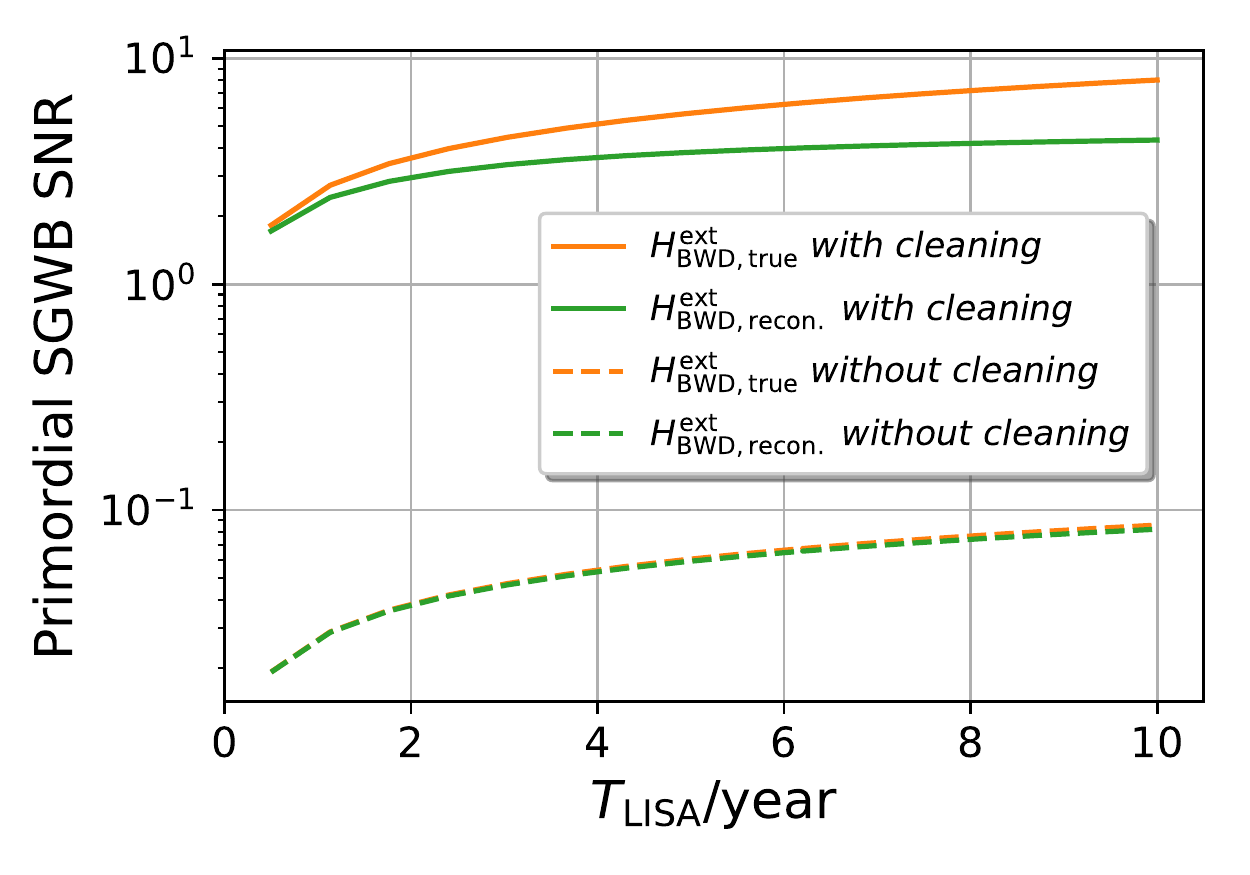}
\includegraphics[scale=0.65]{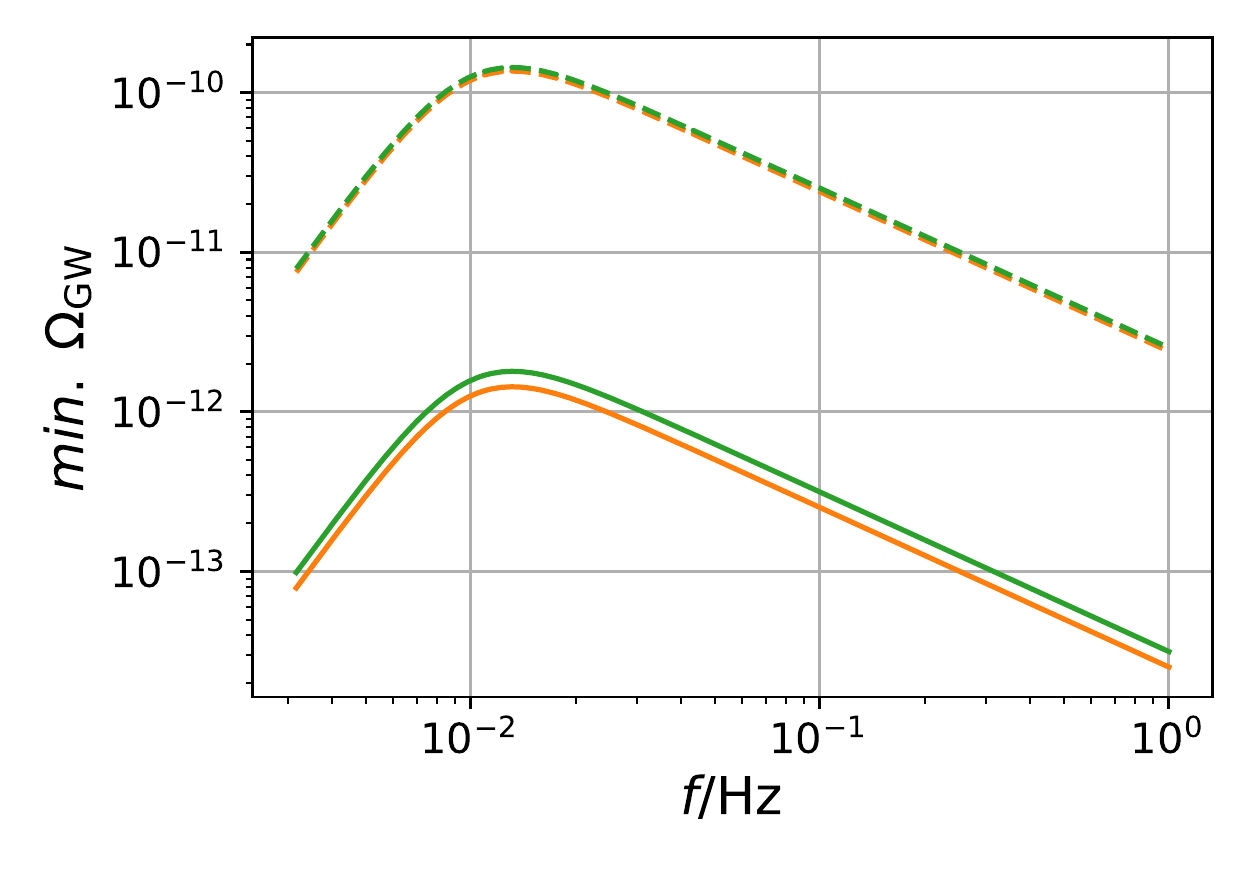}
\caption{\label{fig:sgwb}
Left panel shows SNRs of different estimators for the fiducial primordial SGWB (Eq.[\ref{eq:pex}]), assuming  $\Hext(f)$ is either known (orange, $\sigma_{\rm BWD}=0$) or to be reconstructed (green, $\sigma_{\rm BWD}\neq0$).
Right panels shows the minimum energy $\Omega_{\rm GW}(f)$ of SGWB that can be detected by LISA with $\ge 5\sigma$ confidence level
assuming observation time of 5 years.}
\end{figure}

For illustration purpose,
we consider an example with SGWB generated by bubble collisions during a first order phase transition \cite{Caprini2009}:
\be\label{eq:pex}
H_{\rm P}(f) = \frac{10^{-42}}{1 + (f/0.01 {\rm Hz})^4} \  {\rm Hz}^{-1}\quad
{\rm i.e.,}\quad
\Omega_{\rm GW}(f)= 2.6\times 10^{-12}\frac{(f/0.01 {\rm Hz})^3 } {1 + (f/0.01 {\rm Hz})^4}.
\ee
In the left panel of Fig.~\ref{fig:sgwb}, we show the SNR $U_{\rm P}/\sqrt{\sigma_U^2 + \sigma_{\rm BWD}^2 + \sigma_{\rm A}^2}$
of estimator $\hat U$ (solid lines),
and the SNR $V_{\rm P}/\sqrt{\sigma_V^2 + \sigma_{\rm BWD}^2}$ of estimator $\hat V$ (dashed lines),
as functions of LISA running time, where $\sigma_{\rm BWD}$ is calculated
from the reconstructed shape uncertainty (blue contour in Fig.~\ref{fig:Hext}).
Without the foreground cleaning (estimator $\hat V$), even if we roughly know the spectral shapes of $H_{\rm A}(f)$ and $\Hext(f)$,
the sensitivity on primordial waves is negligible because of their degeneracy with the SGWB.
However, after removing the compact binary foreground (estimator $\hat U$), we find that the sensitivity to the primordial waves is greatly enhanced beyond one order of magnitude. The measurement of compact binary foreground and associated multiband cleaning becomes the critical factor that enables the detection of primordial waves. The reconstruction uncertainty of the extragalactic BWD foreground $\Hext(f)$ mildly degrades the
LISA sensitivity by $5\%\sim 50\%$ depending on the LISA running time.  In the right panel, we show the minimum
energy $\Omega_{\rm GW}(f)$ of SGWB with the same shape of the example SGWB [Eq.~(\ref{eq:pex})] that can be detected by
LISA with $\ge 5\sigma$ confidence level assuming observation time of $5$ years.

The example primordial SGWB above in high-frequency range
\be
\Omega_{\rm GW}(f)\simeq 1.0\times 10^{-15} \left(\frac{25\ {\rm Hz}}{f}\right)\ ,
\ee
might also be constrained directly by ground based detectors. The current best constraint on SGWB from LIGO is $\Omega_{\rm GW}(f= 25\ {\rm Hz})\lesssim 10^{-7}$ (8 orders of magnitude louder than the example signal)
at $95\%$ confidence level \cite{LVC2019b}. Third-generation detectors are expected to be sensitive to primordial SGWB at
the level of $\Omega_{\rm GW}(f= 25\ {\rm Hz})\simeq 10^{-13}$ (2 orders of magnitude louder) assuming 5 years of observation \cite{Regimbau2017}. As shown in Fig.~\ref{fig:sgwb},
the example signal is expected to be detected by LISA
in combination with the multi-band foreground cleaning with SNR=$4\sim 6$ assuming 5 years of LISA observation,
whereas the LISA sensitivity degrades by about two orders of magnitude without the multi-band foreground cleaning
and LISA could be sensitive to the example signal if it was $>20$ times louder.

\section{Discussion}
\label{sec:discussion}
In the main text, we have assumed that all the galactic BWDs with $f>5$ mHz can be completely substracted from the foreground
\cite{Cutler2006}. In fact, there are more complexities in BWD systems that makes the GW emission hard to be accurately modeled, including accretion when the binary separation is small enough and BWDs in three-body systems. In a BWD system, a WD star is expected to fill the Roche lobe if the binary separation is less than $0.1 R_\odot$ \cite{Farmer2003}, when the GW frequency $f_{\rm GW} \gtrsim 0.1$ Hz. Therefore accretion between close BWDs might not important considering the small number of BWDs (at most a few \cite{Lamberts2019}) and the large uncertainty of the BWD foreground in this frequency range (Fig.~\ref{fig:Hext}).
Despite of
many N-body simulations of compact objects in dynamical formation channels in the last a couple years, the fraction of BWDs in three-body systems is unknown and there is no accurate model for their GW emission. If a large fraction of BWDs are confirmed in three-body systems by LISA, more accurate models are necessary for the purpose of foreground cleaning.

For all the estimators,
we have assumed an BBH foreground with a simple
power-law spectral density $H_{\rm A}(f)\propto f^{-7/3}$,
which is true only if binaries have zero eccentricity.
In the case of mildly eccentric binaries, the spectral density has a more complicated frequency dependence,
which deviates from the simple power-law by $\lesssim 50\%$ for BBHs with eccentricity $e \lesssim 0.2$ \cite{Huerta2015, Randall2019}. In addition, highly eccentric binaries formed through direct captures \cite{Rodriguez2018}, as sources for ground-based detectors,   are born at frequencies higher than those spanned by the LISA band.
Therefore it is important to understand the eccentricity distribution of the BBHs
for correctly determining the foreground spectral density in the LISA band.

Currently it is believed that there are two main formation channels:
field binary evolution and dynamical formation in a dense stellar environment.
While BBHs from the former channel are expected to have negligible eccentricities, dynamical formation has the potential
to produce BBHs with high eccentricities. As implied by simulations done in Refs.~\cite{Rodriguez2018,Kremer2019},
a non-negligible fraction ($\sim 28\%\times 20\% = 5.6\%$) of dynamically formed BBHs in dense globular clusters have large eccentricities ($e_{0.01\ {\rm Hz}}\gtrsim 0.1$ or $e_{10\ {\rm Hz}}\gtrsim 10^{-4}$) in the LISA band, and a even larger fraction ($\sim 12\%$) of BBHs are born with
larger eccentricities ($e_{10\ {\rm Hz}}\gtrsim 10^{-3}$ ) and never enter the LISA band.
The latter ones can be readily measured by  CE/ET which can distinguish mergers with eccentricities
$e_{10\ {\rm Hz}}\geq 1.7\times 10^{-4}$ \cite{Lower2018}, so that they will not  affect the foreground estimation.
 As a result, we expect $H_{\rm A}(f)$ to have  $\sim 0.5 \times 5.6\% \times F_{\rm dy}$ deviation from the
circular approximation, where $F_{\rm dy}$ is the fraction of  BBHs born in the dynamical channel.
If dynamical formation is a sub-dominant channel (say, $F_{\rm dy}<0.4$), the deviation is likely at sub-percent level and can be safely ignored for our purpose. On the other hand, the loud BBH events detected by LISA may also provide us information about eccentricity distribution in the LISA band \cite{Breivik2016,Nishizawa2016,Nishizawa2017,Randall2019}.
Last but not least, another space-borne detector Tianqin is designed to be more sensitive to GWs at higher frequencies (than that of LISA)\cite{Lu2019} where the astrophysical foreground is weaker, therefore we expect Tianqin to open another window to probing the primordial SGWB.

In summary, foreground cleaning is a complicated problem and will be a critical factor
for detecting the primordial SGWB, one of the most rewarding scientific goals of space missions.
A glimpse in this paper does not clean up all the complexities and more effort should be devoted into it.
Previous research about detecting various primordial GW
signals without considering astrophysical foregrounds or with idealized foreground cleaning should be reexamined.

\ack
  This research was supported by the
Natural Sciences and Engineering Research Council of
Canada and in part by Perimeter Institute for Theoretical
Physics. Research at
Perimeter Institute is supported in part by the Government
of Canada through the Department of Innovation,
Science and Economic Development Canada and by the
Province of Ontario through the Ministry of Colleges and
Universities.

\appendix

\section{BBH Foreground cleaning by event-based subtraction}
\label{sec:onebyone}

In the main text, we reconstructed the underlying distribution of BBH mergers from all the mergers recorded by the CE
following the statistical approach. Now we explore the foreground measurement of event-to-event approach . Assume the LISA runs from $-T/2$ to $T/2$, and the CE runs from
$-T/2$ to $-T/2+T_{\rm CE}$. For each BBH in the LISA band, we expect to detect its merger after some time $t$ in the CE band,
where $t$ is determined by the GW frequency evolution equation \cite{Cutler1994}
\be\label{eq:t_f}
\frac{dt}{df} = \frac{5}{96\pi^{8/3}} \frac{f^{-11/3}}{[G M_c(1+z)]^{5/3}} \ ,
\ee
and we can add up the constribution to the LISA band foreground from BBHs which merger during the CE running phase,
\be
\hat H_{\rm A}(f)|_{\rm CE} = \frac{1}{T}\sum_i  \left(|h_+(f)|^2+|h_\times(f)|^2 \right)_i \Theta_T^i(f) \ ,\nonumber
\ee
where $i$ runs over all mergers detected by the CE. If the CE runs for
a infinitely long time $T_{\rm CE}\rightarrow \infty$, $\hat H_{\rm A}(f)|_{\rm CE}\rightarrow \hat H_{\rm A}(f)$.
For a finite CE running time, only a fraction of BBHs in the LISA band will evolve into merger phase and the expectation value turns out to be
\be\label{eq:HA_CE_mean}
\begin{aligned}
  \braket{\hat H_{\rm A}(f)|_{\rm CE}}
  &=  \int_{z=z_{\rm min}(f, M_c)}^{z=\infty}\int_{M_{\rm c,min}}^{M_{\rm c,max}}  \sum_A |h_A(f)|^2 \\
  &\times \frac{R(z)}{1+z}dV_c(z) p(M_c) d M_c \ ,
\end{aligned}
\ee
where $z_{\rm min}(f, M_c)$ is determined by Eq.~(\ref{eq:t_f})
\be
T_{\rm CE} =  \frac{15}{768\pi^{8/3}} \frac{f^{-8/3}}{[G M_c(1+z_{\rm min})]^{5/3}} \ ,
\ee
where we have used the fact that the BBH merger frequency is well above the LISA band frequency $f$.

\begin{figure}
\includegraphics[scale=0.65]{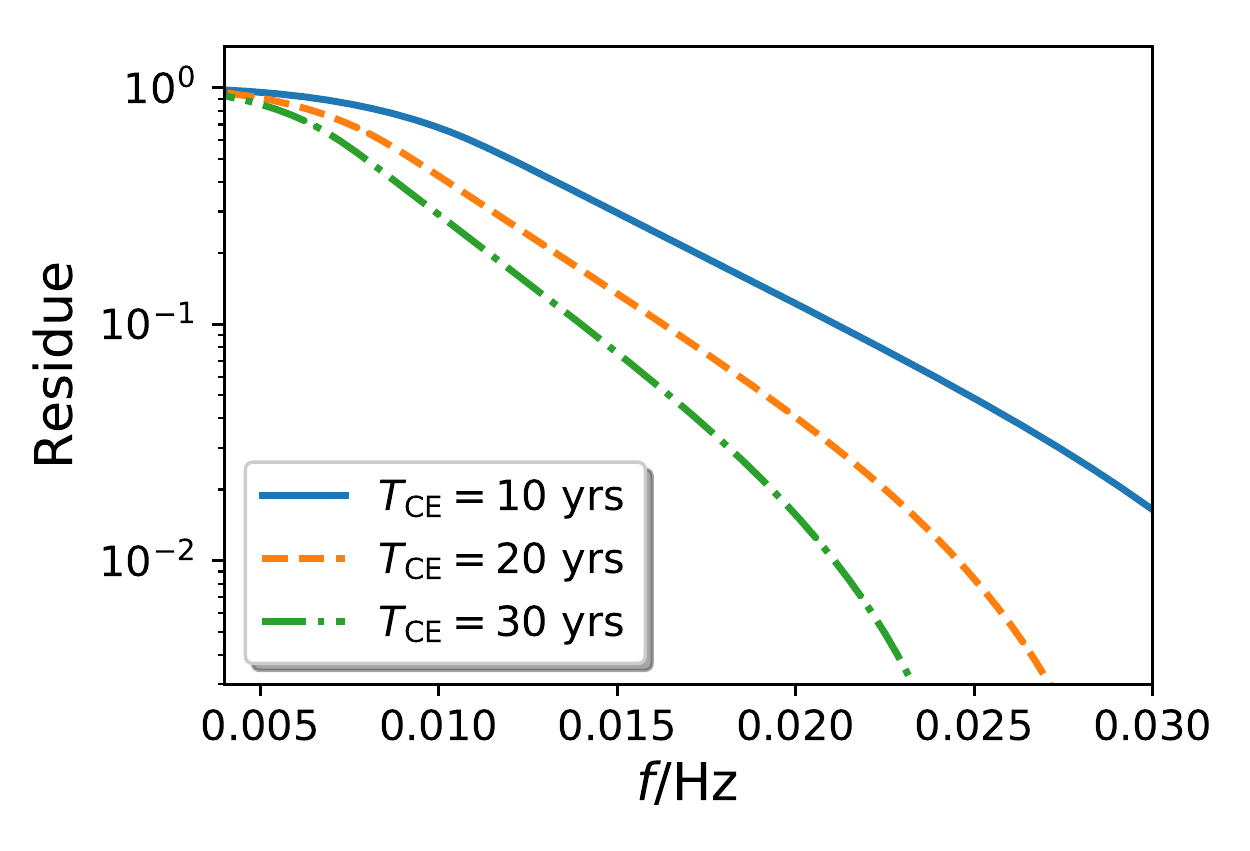}
\caption{\label{fig:oneone} Performance of foreground cleaning with event-based subtraction.}
\end{figure}

In Fig.~\ref{fig:oneone}, we show the residue fraction  after the event-to-event cleaning
$1-\braket{\hat H_{\rm A}(f)|_{\rm CE}}/\braket{\hat H_{\rm A}(f)}$  given a finite running CE time ($T_{\rm CE} = 10, 20, 30$ years). It takes longer time for binaries of lower frequency to merge, therefore lower-frequency  foreground cleaning is slower. With the CE running for $10$ years, the astrophysical foreground can be
cleaned to percent level for $f \gtrsim 0.03$ Hz.
Therefore the foreground cleaning of distribution-based approach
is more efficient than that of event-based approach.

\section{Stochastic GWs from Binary White Dwarfs}
\label{sec:Hext}
Similar to BBHs,  the spectral density of the Binary White Dwarf (BWD)
foreground averaged over time period $[-T/2, T/2]$ is also
\be
\hat H_{\rm BWD}(f) = \frac{1}{T}\sum_i  \left(|h_+(f)|^2+|h_\times(f)|^2 \right)_i \Theta_T^i(f) \ ,
\ee
where $f$ is a frequency in the LISA band. Unlike BBHs, BWDs merge in the LISA band, and this results in
some frequency dependence of the expectation value  $\braket{\frac{1}{T}\sum_i \Theta_T^i(f)}$.
This extra frequency dependence drives $ H_{\rm BWD}(f)$ off the power law $f^{-\frac{7}{3}}$
and it is hard to calculate from first principle.

According to simulations performed in Ref.~\cite{Lamberts2019}, all the high-frequency (say $f> 5$ mHz) galactic BWDs are expected
to be resolved in the LISA mission time. From these BWDs, we can construct a normalized spectral density,
\be
H_{D}(f) \propto \frac{1}{T}\sum_i  \left(\sum_A|h_A(f)|^2 D_{\rm L, eff}^2\right)_i \Theta_T^i(f) \ ,
\ee
where all the effective distance information has been removed,
$D_{\rm L,eff} = D_{\rm L}/\sqrt{\cos^2\imath + (\frac{1+\cos^2\imath}{2})^2}$. The normalized spectral density
$H_D(f)$ is actually a statistical property of the galactic BWDs and can be used to reconstruct the extragalactic
BWD foreground if the population of galactic BWDs does not deviate significantly from average
extraglatic BWDs. The spectral density of extragalactic BWD foreground is then
calculated as
\be\label{eq:Hext}
\Hext(f) \propto \int_0^{V_c(z_*)}  \frac{ H_{D}(f(1+z))}{D_{\rm L}^2(z)} \frac{\mathcal F(z)}{1+z} dV_c(z)\ ,
\ee
where $\mathcal F(z)$ is the formation rate of high-frequency BWDs. Though the accurate form of $\mathcal F(z)$
is unknown, as shown in Fig.~\ref{fig:Hext}, it has little influence on the shape of $\Hext(f)$.

There are two major sources of the uncertainty of $\Hext(f)$ reconstruction from resolvable galactic BWDs,
one is statistic uncertainty due to limited number of galactic BWDs and the other is the uncertainty in the formation rate
$\mathcal F(z)$.
Following \cite{Lamberts2019}, we assume the total number of resolvable high-frequency galactic BWDs is $10^4$,
and these BWDs roughly satisfy a power-law distribution $dN/df\propto f^{-4.8}$ in the frequency range $[5, 120]$ mHz.
For the chirp mass, we assume a Gaussian distribution with mean value $0.4 M_\odot$ and standard deviation $0.1 M_\odot$.
To examine how much uncertainty in $\Hext(f)$ is introduced by the uncertainty of $\mathcal F(z)$, we consider two
$\mathcal F(z)$ functions: one is the star formation rate \cite{Cole2001}, and the other is a constant.
For each $F(z)$, we simulate 100 realizations of resolved galactic BWDs, for each of which we fit $\log H_D-\log f$ with
a 2nd-degree polynomial and calculate $\Hext(f)$ from Eq.~(\ref{eq:Hext}).
The true spectral density $\Hext(f)$ along with
the reconstruction uncertainties are shown in Fig.~\ref{fig:Hext},
which clearly shows the shape of $\Hext(f)$ is insensitive to $\mathcal F(z)$
and the reconstruction uncertainty is dominated by the statistic uncertainty.

The reconstruction uncertainty is small ($\sim 1\%$)  at low frequencies where a large number of BWDs reside, and
is much larger ($\sim 200\%$) at high frequencies.  However, this method relies on the assumption that
the population of galactic BWDs does not deviate significantly from average extragalactic BWDs.
This assumption may be tested with population synthesis studies and further examined by comparing
the observationally reconstructed background with the prediction in Ref.~\cite{Farmer2003}.

\section{One detector, two channels}
\label{sec:onedetector}

\begin{figure}
\includegraphics[scale=0.65]{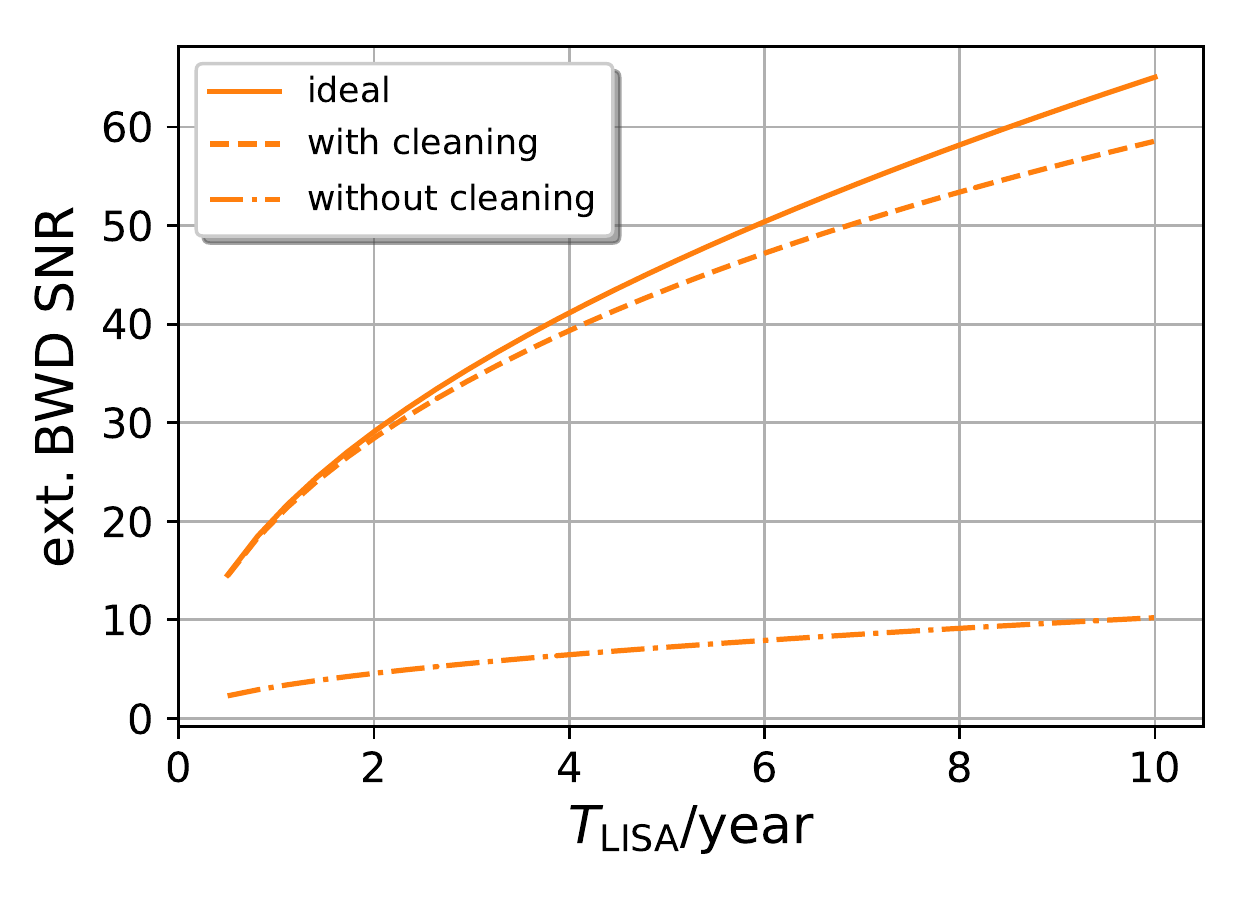}
\caption{\label{fig:2ch} Measuring the extragalactic BWD foreground using two channels of a single LISA detector. }
\end{figure}

In the main text, we have outlined the foreground cleaning with KDE reconstruction assuming two LISA detectors for convenience,
where the stochastic GWs can be separated from detector noise by cross-correlating the two detector outputs.
For the proposed LISA mission, there will be a single detector and we may detect the SGWB by utilizing the ``null"  channel which is blind to GW signals for detector noise calibration, in combination with the normal Michelson (m) channel \cite{Romano2017}.
In reality, a good approximation to the ideal null channel is the symmetrized Sagnac (ss) channel whose response function
$\mathcal R_{\rm ss}(f)$ is much smaller than that of the Michelson channel $\mathcal R_{\rm m}(f)$ especially in the lower frequency range \cite{Armstrong1999, Tinto2000, Cornish2001b, Hogan2001}.
In combination with the outputs of the two channels, it is natural to write the SGWB estimator as
\be
\begin{aligned}
  \hat X &=  \int_{-\infty}^{\infty} df \int_{-\infty}^{\infty} df' \delta_T(f-f')   \\
  &\times \left[s^*_\m(f) s_\m(f') - W(f) s^*_\rss(f) s_\rss(f')\right] Q(f)\ ,
\end{aligned}
\ee
where $W(f) = P_{\m}(f)/P_{\rss}(f)$, with $P_{\rss}(f)$ and $P_{\m}(f)$ being
the detector noise spectral density of two channels \cite{Robson2019}. The mean value and variance are
\be
\begin{aligned}
  \braket{\hat X}
  &= T\int_0^\infty H(f)\left[\mathcal R_\m(f)-W(f)\mathcal R_\rss(f)\right]Q(f) df, \\
  \sigma_X^2 &\approx 2T\int_0^\infty \left[ P^2_{\m}(f) -W(f) |P_{\m,\rss}(f)|^2\right] |Q(f)|^2 df \ . \nonumber
\end{aligned}
\ee
where $P_{\m,\rss}(f)$ is cross power of noises in the two channels.
In the similar way, we can write the estimators $\hat Y$ and $\hat Z$ as in the case of two  LISA detectors.
We show the SNRs of different estimators for the extragalactic BWD foreground in Fig.~\ref{fig:2ch}.
As in the two-detectors case, the BBH foreground cleaning increases the LISA sensitivity by a factor $4\sim 7$.
Compared with the case of two detectors, the SNRs of different estimators here turn out to be smaller by a factor $\sim 2$.

\section*{References}
\bibliographystyle{iopart-num}
\bibliography{ms}
\end{document}